\documentclass[twocolumn,final,times]{elsarticle}
\usepackage{graphicx,amssymb,amsmath}
\journal{Physica B}

\begin{document}

\begin{frontmatter}

\title{Giant magnetocaloric effect, magnetization plateaux and jumps of the regular Ising polyhedra\tnoteref{grant}}
\tnotetext[grant]{This work was financially supported by the grant of the Slovak Research and Development Agency under the contract No. APVV-0132-11.}
\author[UFV]{Jozef Stre\v{c}ka\corref{coraut}} 
\cortext[coraut]{Corresponding author}
\ead{jozef.strecka@upjs.sk}
\author[UFV]{Katar\'ina Kar\v{l}ov\'a}
\author[UMV]{Tom\'a\v{s} Madaras}
\address[UFV]{Institute of Physics, Faculty of Science, P. J. \v{S}af\'{a}rik University, Park Angelinum 9, 040 01 Ko\v{s}ice, Slovak Republic}
\address[UMV]{Institute of Mathematics, Faculty of Science, P. J. \v{S}af\'{a}rik University, Jesenn\'a 5, 040 01 Ko\v{s}ice, Slovak Republic}

\begin{abstract}
Magnetization process and adiabatic demagnetization of the antiferromagnetic Ising spin clusters with the shape of regular polyhedra (Platonic solids) 
are exactly examined within the framework of a simple graph-theoretical approach. While the Ising cube as the only unfrustrated (bipartite) spin cluster shows 
just one trivial plateau at zero magnetization, the other regular Ising polyhedra (tetrahedron, octahedron, icosahedron and dodecahedron) additionally 
display either one or two intermediate plateaux at fractional values of the saturation magnetization. The nature of highly degenerate ground states 
emergent at intermediate plateaux owing to a geometric frustration is clarified. It is evidenced that the regular Ising polyhedra exhibit a giant magnetocaloric effect 
in a vicinity of magnetization jumps, whereas the Ising octahedron and dodecahedron belong to the most prominent geometrically frustrated spin clusters that
enable an efficient low-temperature refrigeration by the process of adiabatic demagnetization.   
\end{abstract}


\end{frontmatter}

\section{Introduction}

A recent progress in a targeted design of molecular nanomagnets, which afford paradigmatic examples of assemblies of a finite number of interacting spin centers, has triggered a considerable interest to explore a magnetic behavior of small spin clusters magnetically isolated from the environment \cite{carl86,kahn93,gatt06,furr13}. Despite their simplicity, small spin clusters may still exhibit favorable magnetic properties such as the quantum tunneling of magnetization and slow spin relaxation observable in single-molecule magnets, which have an immense application potential in a development of new resources suitable for quantum information processing and/or novel generation of high-density storage devices \cite{leue01,troi05,frie10}. In addition, the molecular nanomagnets being composed of small spin clusters are ideal for an experimental testing of the limitations of physical theories especially when physical properties of relevant model systems can be calculated without any uncontrolled approximation.

Magnetic properties of small spin clusters are predominantly determined by the nature of magnetic interactions between the spin centers in a cluster, whereas the Heisenberg superexchange coupling is usually the most dominant interaction term that basically influences characteristic features of molecular nanomagnets \cite{carl86,kahn93,gatt06,furr13}. A lot of attention has been therefore paid to the study of antiferromagnetic Heisenberg spin clusters, which may exhibit striking intermediate plateaux in low-temperature magnetization curves that often macroscopically manifest highly non-trivial quantum ground states \cite{park00,schn01,schm05,schr05a,schr05b,kons05,schn06,kons07,rous08,schn09,kons09,saho11,umme13}. An influence of the exchange, dipolar and single-ion anisotropy on the low-temperature magnetization process of the antiferromagnetic Heisenberg spin clusters has attracted much less attention so far \cite{shap02,efre06,efre08,huch11}. For this reason, it appears worthwhile to investigate the low-temperature magnetization process of antiferromagnetic Ising spin clusters, which are also capable of displaying several intermediate magnetization plateaux on assumption that a spin cluster is geometrically frustrated \cite{lieb86,diep04}. 

The main goal of the present work is to examine the magnetization process and adiabatic demagnetization of the antiferromagnetic Ising spin clusters with the shape of regular polyhedra (Platonic solids), which surprisingly seem not to be dealt with previously. The zero-field thermodynamics of the regular Ising polyhedra with the uniform interaction have been explored in some detail by Syozi \cite{syoz55} and Fisher \cite{fish59} within the framework of dual, decoration-iteration and star-triangle transformations, while the regular Ising polyhedra with the mixed ferromagnetic and antiferromagnetic couplings have been examined in relation with the spin-glass physics \cite{voge93,vald00,lebr04}. It is noteworthy that the competition between the antiferromagnetic order, spin frustration and magnetic field has been studied by Viitala and co-workers by considering antiferromagnetic Ising spin clusters with several cluster geometries as for instance octahedron, body centered icosahedron and cubooctahedron \cite{vita97a,vita97b}. More recently, an exact enumeration of states has been employed in order to calculate numerically the residual entropy, magnetocaloric properties and magnetization process of various geometrically frustrated Ising spin clusters composed of triangular units \cite{zuko14,zuko15}.   

The organization of this paper is follows. Exact results for the partition function, free energy and magnetization of the regular Ising polyhedra are derived in Section \ref{method} within the framework of a graph-theoretical approach. The most interesting results obtained for the magnetization process and adiabatic demagnetization of the regular Ising polyhedra are presented and discussed in detail in Section \ref{result}. Finally, some conclusions and future outlooks are drawn in Section \ref{conclusion}.

\section{Regular Ising polyhedra}
\label{method}

\begin{figure*}[t]
\begin{center}
\includegraphics[width=0.8\textwidth]{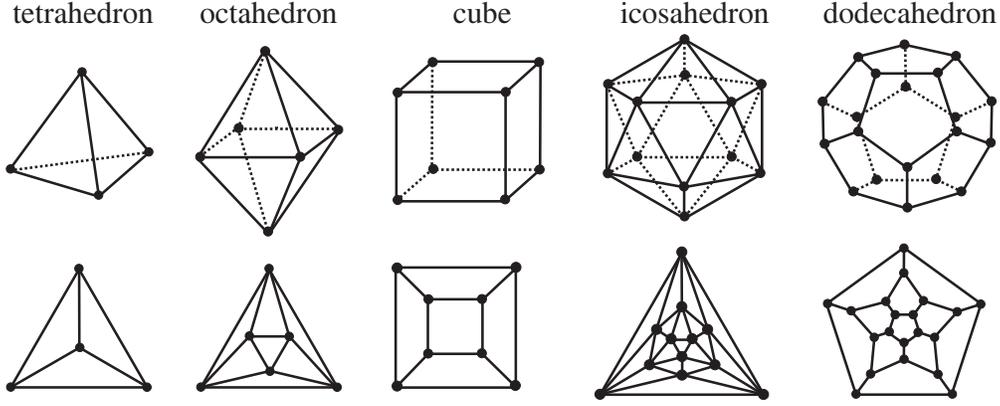}
\end{center}
\vspace{-0.4cm}
\caption{The Ising spin clusters with the shape of five regular polyhedra. The upper panel displays a steric arrangement of the regular polyhedra, 
while the lower panel shows the equivalent planar lattice graphs.}
\label{fig1}
\end{figure*}

Let us consider the Ising spin clusters with the geometry of five regular polyhedra (tetrahedron, octahedron, cube, icosahedron and dodecahedron) as depicted in Fig.~\ref{fig1}. 
The regular Ising polyhedra can be defined through the following Hamiltonian
\begin{equation}
H = J \sum_{\langle i,j \rangle}^{N_b} S_i S_j - h \sum_{i=1}^{N} S_i, 
\label{ham}
\end{equation}
where $S_i = \pm 1$ represents the Ising spin placed at $i$th vertex of a regular polyhedron, the first summation accounts for the antiferromagnetic Ising-type exchange interaction $J>0$ between adjacent spins, the second summation accounts for the Zeeman's energy of individual magnetic moments in the external magnetic field $h>0$ and finally, $N$ $(N_b)$ stands for the total number of vertices (edges) of a regular polyhedron that is simultaneously equal to the total number of spins (bonds). It is noteworthy that the considered Ising spin clusters are sufficiently small finite-size systems in order to calculate all physical quantities of interest quite rigorously. In the following, we will calculate the partition function, free energy and magnetization of the regular Ising polyhedra within a relatively simple graph-theoretical approach.

The canonical partition function of the regular Ising polyhedra is defined through the relation
\begin{equation}
Z = \sum_{\{S_i\}} \exp(-\beta H), 
\label{pf}
\end{equation}
where $\beta = 1/(k_{\rm B} T)$, $k_{\rm B}$ is Boltzmann's constant, $T$ is the absolute temperature and the summation $\sum_{\{S_i\}}$ is carried out over all possible configurations of a set of the Ising spins $\{S_i\}$. While the interaction part of the Hamiltonian (the first term in Eq.~(\ref{ham})) depends on many specific details of a particular spin configuration, the Zeeman's energy (the second term in Eq.~(\ref{ham})) solely depends on the total spin $S_T = \sum_{i=1}^{N} S_i$ and is independent of any other specific detail of a given spin configuration. It is therefore quite convenient to sort available spin configurations according to the total spin $S_T$, because only the lowest-energy spin configuration from a given set of microstates with the same total spin $S_T$ may eventually become a ground state. In addition, it is sufficient to consider only spin configurations with non-negative values of the total spin $S_T \geq 0$, since the exchange energy is invariant under the reversal of all Ising spins ($S_i \rightarrow - S_i$ for each $i$) and the Zeeman's energy merely changes its sign. 

To proceed further with a calculation, let us exploit the one-to-one correspondence between the Ising spins $S_i = \pm 1$ and the newly defined two-valued variables $n_i = 0,1$ established 
through the formula
\begin{equation}
S_i = 1 - 2 n_i, 
\label{map}
\end{equation}
which connects the 'up' spin state $S_i = 1$ to the value $n_i = 0$ and the 'down' spin state $S_i = -1$ to the other value $n_i = 1$. The Hamiltonian (\ref{ham}) can be consequently 
rewritten in terms of newly defined two-valued variables $n_i$ to the following form 
\begin{eqnarray}
H \!\!\!&=&\!\!\! J \left[N_b - 2 \sum_{\langle i,j \rangle}^{N_b} \left(n_i + n_j - 2 n_i n_j \right) \right] \nonumber \\
  \!\!\!&-&\!\!\! h \left(N - 2 \sum_{i=1}^{N} n_i \right). 
\label{hamn}
\end{eqnarray}
It is quite evident from Eq.~(\ref{hamn}) that the exchange energy $N_b J$ of the fully aligned spin configuration can be lowered by an amount $-2J$ per each unequally oriented pair of adjacent Ising spins, but each Ising spin flipped in opposite to the external magnetic field $S_i = -1$ $ (n_i = 1)$ simultaneously raises the Zeeman's energy $-N h$ by an amount $2h$. Apparently, the lowest-energy spin configuration (ground state) at a given magnetic field must be a reasonable compromise that would minimize a sum of both parts of the overall energy. 

The alternative form of the Hamiltonian (\ref{hamn}) also suggest an interesting mapping correspondence between the summation over all available configurations of the Ising spins and a relatively simple graph-theoretical counting problem. When the Ising spins oriented in opposite to the magnetic field $S_i=-1$ ($n_i=1$) are identified as vertices of the equivalent planar graph (see lower panel in Fig.~\ref{fig1}), the combinatorial problem of finding all possible spin configurations becomes equivalent to the problem of finding all induced subgraphs of a given planar graph. As a matter of fact, the total number of the flipped Ising spins ($S_i=-1$, $n_i=1$) then directly equals to the total number of vertices $n_t = \sum_{i=1}^{N} n_i$ of the corresponding induced subgraph and, respectively, the total number of unequally oriented adjacent spin pairs $\sum_{\langle i,j \rangle}^{N_h} \left(n_i + n_j - 2 n_i n_j \right)$ equals to a sum over complementary degrees of all vertices present in a given induced subgraph $\tilde{d}_t = \sum_{i=1}^{N} n_i \tilde{d}_i = \sum_{i=1}^{N} n_i (d - d_i)$ ($d$ is degree of a vertex in a full graph and $d_i$ is degree of the $i$th vertex in a given induced subgraph). Hence, it follows that the overall energy of each individual spin configuration can be straightforwardly calculated from the total number of vertices $n_t$ and the sum of all complementary degrees $\tilde{d}_t$ of the corresponding induced subgraph according to the formula 
\begin{equation}
E = J (N_h - 2 \tilde{d}_t) - h (N - 2 n_t). 
\label{enc}
\end{equation}
It is worthwhile to remark that the graphs corresponding to the regular Ising polyhedra are vertex-transitive, which substantially simplifies the relevant calculation of complementary degrees of vertices in respective induced subgraphs. 

\begin{figure*}
\vspace{-1cm}
\begin{center}
\includegraphics[width=0.80\textwidth]{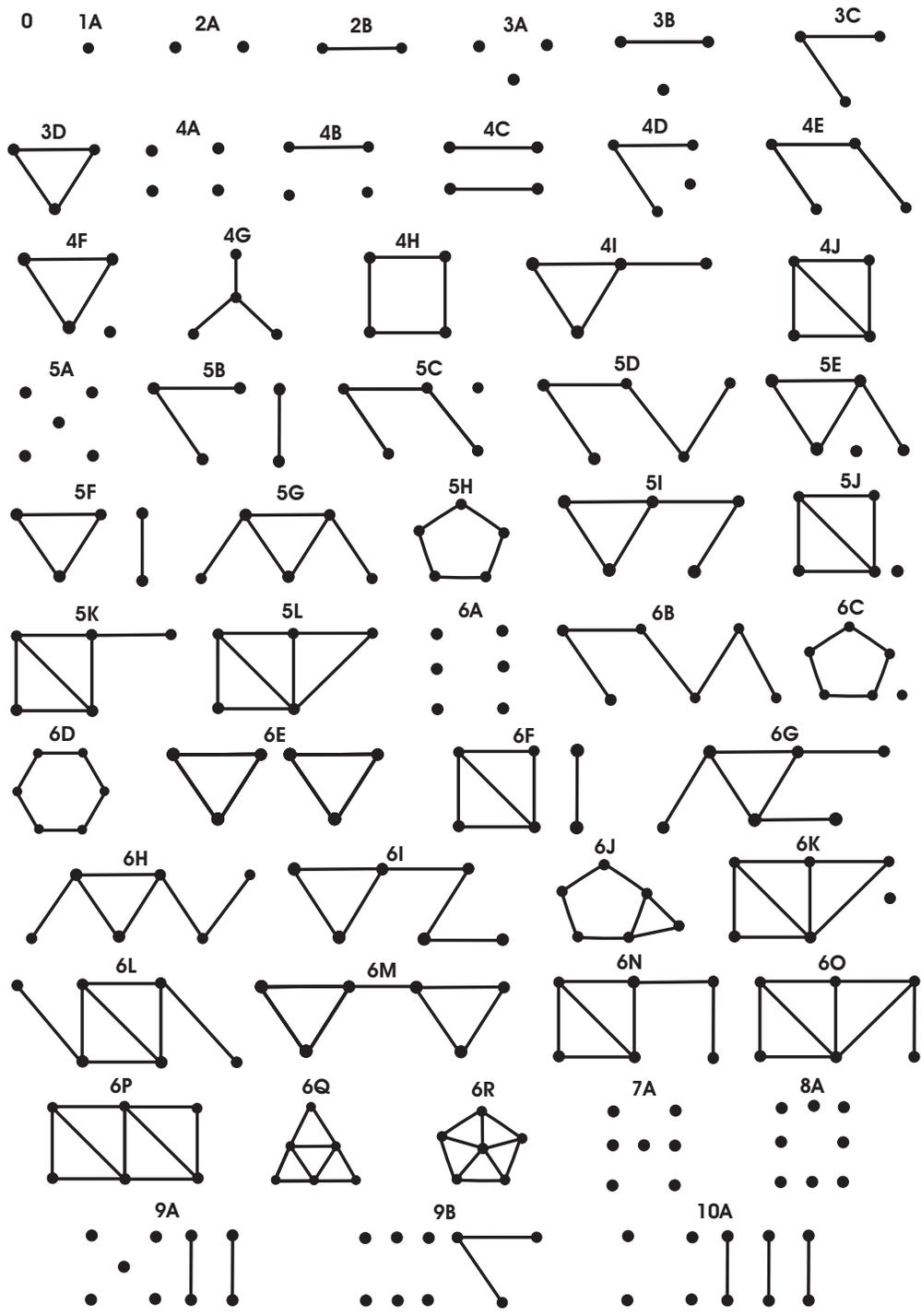}
\end{center}
\caption{A schematic representation of the induced subgraphs, which correspond to all possible spin configurations of the Ising tetrahedron, octahedron, cube and icosahedron, as well as, 
the lowest-energy spin configurations of the Ising dodecahedron.}
\label{fig2}
\end{figure*}

Let us illustrate the aforedescribed calculation procedure on particular case of the regular Ising polyhedra starting from the simplest example, which is the Ising tetrahedron (see Tab.~\ref{TT}). When all four spins are in the 'up' spin state $S_i=1$ ($n_i=0$) one obtains a fully aligned spin configuration of the Ising tetrahedron with the total spin $S_T=4$, which corresponds according to the established mapping equivalence to a very special subgraph $0$ (empty graph) with a zero vertex set. All other microstates of the Ising tetrahedron with $S_T \geq 0$ can be then obtained from this fully aligned spin configuration by turning a desired number of the Ising spins to their 'down' spin state, whereas it is sufficient to flip no more than a half of the total number of spins owing to the invariance of the exchange energy with respect to the inter-change $S_i \to -S_i$. Bearing this in mind, only two additional spin configurations of the Ising tetrahedron should be taken into account with either one or two inverted spins, whereas the former spin configuration with $S_T=2$ corresponds to an induced subgraph containing an isolated vertex (the subgraph 1A in Fig.~\ref{fig2}) and the latter spin configuration with $S_T=0$ to an induced subgraph involving two vertices incident to a common edge (the subgraph 2B in Fig.~\ref{fig2}). It is worthwhile to remember that the overall energy of spin configurations with $S_T<0$ can be obtained from that ones with $S_T>0$ simply by inverting the sign of Zeeman's term (the exchange energy is invariant under this reversal).

\begin{table}
\vspace{-1.0cm}
\caption{Spin configurations of the Ising tetrahedron classified according to the total spin $S_{T} \geq 0$, the total number of flipped Ising spins ($n_t$), the degeneracy (deg), the total number of unequally oriented adjacent spin pairs ($\tilde{d}_t$), the overall energy and the corresponding induced subgraph (see Fig.~\ref{fig2} for schematic representation of induced subgraphs). The values $n_t$ and $\tilde{d}_t$ coincide with the total number of vertices in a given subgraph and the sum of their complementary degrees, respectively.}
\vspace{-0.2cm}
\begin{center}
\begin{tabular}{|c|c|c|c|c|c|}
\hline $ S_{T} $ & $ n_t $ & deg & $ \tilde{d}_t $ & energy & subgraph \\ 
\hline \textbf{4} & \textbf{0} & \textbf{1} & \textbf{0} & $\boldsymbol{6J - 4h}$ & \textbf{0} \\ 
\hline \textbf{2} & \textbf{1} & \textbf{4} & \textbf{3} & $\boldsymbol{-2h}$ & \textbf{1A} \\ 
\hline \textbf{0} & \textbf{2} & \textbf{6} & \textbf{4} & $\boldsymbol{-2J}$ & \textbf{2B} \\ 
\hline 
\end{tabular} 
\end{center}
\label{TT}
\end{table}

The aforedescribed calculation procedure can be straightforwardly adapted also to the larger Ising spin clusters, so let us merely emphasize the most essential differences to emerge with increasing a spin-cluster size. It can be already understood from the spin configurations of the Ising octahedron listed in Tab.~\ref{TO} that there may appear a few different spin configurations with the same value of the total spin $S_T$, but which need not have the same configurational energy. For instance, two overturned spins in the Ising octahedron may be or not nearest neighbours with respect to each other and hence, the relevant spin configurations necessarily correspond to two different induced subgraphs: one is being composed of two vertices incident to a common edge (the subgraph 2B in Fig.~\ref{fig2}) and the other one consists of two isolated vertices (the subgraph 2A in Fig.~\ref{fig2}). It is quite clear that the latter spin configuration has lower energy due to a greater total number of unequally oriented adjacent spin pairs (i.e. greater sum of complementary degrees of vertices in a given induced subgraph) in comparison with the former spin configuration, which can be consequently excluded from possible candidates for a ground state. A similar finding concerns with two spin configurations of the Ising octahedron with three reversed spins, which may be or not all nearest neighbours with respect to each other (see the subgraphs 3C and 3D in Fig.~\ref{fig2}). For easy orientation, the spin configurations that represent actual ground state over a finite range of magnetic fields are highlighted bold in Tab.~\ref{TT}--\ref{TI}. 
  
\begin{table}
\vspace{-1.0cm}
\caption{The Ising octahedron (see Tab.\ref{TT} for notation).}
\vspace{-0.2cm}
\begin{center}
\begin{tabular}{|c|c|c|c|c|c|}
\hline $ S_{T} $ & $ n_t $ & deg & $ \tilde{d}_t $ & energy & subgraph \\  
\hline \textbf{6} & \textbf{0} & \textbf{1} & \textbf{0} & $\boldsymbol{ 12J - 6h }$ & \textbf{0} \\ 
\hline 4 & 1 & 6 & 4 & $ 4J - 4h $ & 1A \\ 
\hline \textbf{2} & \textbf{2 }& \textbf{3} & \textbf{8} &$\boldsymbol{  -4J - 2h }$ & \textbf{2A} \\ 
\hline 2 & 2 & 12 & 6 & $ -2h $ & 2B \\
\hline 0 & 3 & 12 & 8 & $ -4J $ & 3C \\ 
\hline 0 & 3 & 8 & 6 & $ 0 $ & 3D \\ 
\hline 
\end{tabular} 
\end{center}
\label{TO}
\end{table}

All spin configurations of the Ising cube with the non-negative total spin $S_T \geq 0$ are listed in Tab.~\ref{TK}. As one can see, a greater capability of the Ising cube in generating the spin configurations with the same total spin $S_T$ can be responsible for an accidental degeneracy of a few different microstates with the identical configurational energy. Another interesting observation is that the lowest-energy spin configurations from a given set of microstates with the same value of the total spin $S_T$ are the ones, in which occupation of the nearest-neighbour vertices by any pair of inverted spins is forbidden because the overall energy is always minimized by the induced subgraphs entirely composed of a set of isolated vertices (see the subgraphs 1A, 2A, 3A and 4A in Fig.~\ref{fig2}). This special property is a direct consequence of the fact that the Ising cube (as the only regular polyhedron) is a bipartite spin cluster. 

\begin{table}
\vspace{-1.0cm}
\caption{The Ising cube (see Tab.\ref{TT} for notation).}
\vspace{-0.2cm}
\begin{center}
\begin{tabular}{|c|c|c|c|c|c|}
\hline $ S_{T} $ & $ n_t $ & deg & $ \tilde{d}_t $ & energy & subgraph \\  
\hline \textbf{8} & \textbf{0} & \textbf{1} & \textbf{0} & $\boldsymbol{12J-8h}$ & \textbf{0} \\ 
\hline 6 & 1 & 8 & 3 & $6J-6h$ & 1A \\ 
\hline 4 & 2 & 16 & 6 & $-4h$ & 2A \\ 
\hline 4 & 2 & 12 & 4 & $4J-4h $ & 2B \\ 
\hline 2 & 3 & 8 & 9 & $-6J-2h $ & 3A \\ 
\hline 2 & 3 & 24 & 7 & $-2J-2h$ & 3B \\ 
\hline 2 & 3 & 24 & 5 & $2J-2h$& 3C \\ 
\hline \textbf{0} & \textbf{4} & \textbf{2} & \textbf{12} & $\boldsymbol{-12J}$ & \textbf{4A} \\ 
\hline 0 & 4 & 6 & 8 & $-4J $ & 4C \\ 
\hline 0 & 4 & 24 & 8 & $-4J $ & 4D \\ 
\hline 0 & 4 & 24 & 6 & $0$ & 4E \\ 
\hline 0 & 4 & 8 & 6 & $0$ & 4G \\
\hline 0 & 4 & 6 & 4 & $4J$ & 4H \\  
\hline 
\end{tabular} 
\end{center}
\label{TK}
\end{table}

Let us also shortly comment on a striking accidental degeneracy of the lowest-energy spin configurations of the Ising icosahedron, which constitute the highly degenerate ground-state manifold with the total spin $S_T=0$ (see Tab.~\ref{TI}). It actually turns out that the true ground state of the Ising icosahedron at small enough magnetic fields is the highly degenerate ground-state manifold, in which six overturned spins either form a path of six vertices (the subgraph 6B in Fig.~\ref{fig2}) or one isolated vertex plus a cycle with five vertices (the subgraph 6C in Fig.~\ref{fig2}). The other highly degenerate lowest-energy spin configurations of the Ising icosahedron can be found among the microstates with five inverted spins (i.e. the total spin $S_T=2$), which either constitute two disjoint paths composed of two and three vertices (the subgraph 5B in Fig.~\ref{fig2}) or an isolated vertex supplemented with a path graph of four vertices (the subgraph 5C in Fig.~\ref{fig2}). It is worthwhile to remark, however, that the lowest-energy spin configurations with the total spin $S_T=2$ do not represent true ground states unlike the spin configurations with the total spin $S_T=0$.

\begin{table}
\vspace{-1.0cm}
\caption{The Ising icosahedron (see Tab.\ref{TT} for notation).}
\vspace{-0.2cm}
\begin{center}
\begin{tabular}{|c|c|c|c|c|c|}
\hline $ S_{T} $ & $ n_t $ & deg & $ \tilde{d}_t $ & energy & subgraph \\ 
\hline \textbf{12} & \textbf{0} & \textbf{1} & \textbf{0} & $\boldsymbol{30J-12h}$ & \textbf{0} \\ 
\hline 10 & 1 & 12 & 5 & $20J-10h$ & 1A \\ 
\hline 8 & 2 & 36 & 10 & $10J-8h$ & 2A \\ 
\hline 8 & 2 & 30 & 8 & $14J-8h$ & 2B \\ 
\hline \textbf{6} & \textbf{3} & \textbf{20} & \textbf{15} & $\boldsymbol{-6h}$ & \textbf{3A} \\ 
\hline 6 & 3 & 120 & 13 & $4J-6h$ & 3B \\ 
\hline 6 & 3 & 60 & 11 & $8J-6h$ & 3C \\ 
\hline 6 & 3 & 20 & 9 & $12J-6h$ & 3D \\
\hline \textbf{4} & \textbf{4} & \textbf{30} & \textbf{18} & $\boldsymbol{-6J-4h}$ & \textbf{4B} \\  
\hline 4 & 4 & 75 & 16 & $-2J-4h$ & 4C \\ 
\hline 4 & 4 & 120 & 16 & $-2J-4h$ & 4D \\ 
\hline 4 & 4 & 120 & 14 & $2J-4h$ & 4E \\ 
\hline 4 & 4 & 60 & 14 & $2J-4h$ & 4F \\ 
\hline 4 & 4 & 60 & 12 & $6J-4h$ & 4I \\ 
\hline 4 & 4 & 30 & 10 & $10J-4h$ & 4J \\ 
\hline 2 & 5 & 60 & 19 & $-8J-2h$ & 5B \\ 
\hline 2 & 5 & 60 & 19 & $-8J-2h$ & 5C \\ 
\hline 2 & 5 & 180 & 17 & $-4J-2h$ & 5D \\ 
\hline 2 & 5 & 60 & 17 & $-4J-2h$ & 5E \\ 
\hline 2 & 5 & 60 & 17 & $-4J-2h$ & 5F \\ 
\hline 2 & 5 & 60 & 15 & $-2h$ & 5G \\ 
\hline 2 & 5 & 12 & 15 & $-2h$ & 5H \\ 
\hline 2 & 5 & 120 & 15 & $-2h$ & 5I \\ 
\hline 2 & 5 & 60 & 15 & $-2h$ & 5J \\ 
\hline 2 & 5 & 60 & 13 & $4J-2h$ & 5K \\ 
\hline 2 & 5 & 60 & 11 & $8J-2h$ & 5L \\ 
\hline \textbf{0} & \textbf{6} & \textbf{60} & \textbf{20} & $\boldsymbol{-10J}$ & \textbf{6B} \\ 
\hline \textbf{0} & \textbf{6} & \textbf{12} & \textbf{20} & $\boldsymbol{-10J}$ & \textbf{6C} \\ 
\hline 0 & 6 & 40 & 18 & $-6J$ & 6D \\ 
\hline 0 & 6 & 10 & 18 & $-6J$ & 6E \\  
\hline 0 & 6 & 30 & 18 & $-6J$ & 6F \\ 
\hline 0 & 6 & 20 & 18 & $-6J$ & 6G \\ 
\hline 0 & 6 & 120 & 18 & $-6J$ & 6H \\
\hline 0 & 6 & 120 & 18 & $-6J$ & 6I \\ 
\hline 0 & 6 & 60 & 16 & $-2J$ & 6J \\ 
\hline 0 & 6 & 60 & 16 & $-2J$ & 6K \\ 
\hline 0 & 6 & 30 & 16 & $-2J$ & 6L \\ 
\hline 0 & 6 & 30 & 16 & $-2J$ & 6M \\ 
\hline 0 & 6 & 120 & 16 & $-2J$ & 6N \\  
\hline 0 & 6 & 120 & 14 & $2J $ & 6O \\ 
\hline 0 & 6 & 60 & 12 & $6J$ & 6P \\ 
\hline 0 & 6 & 20 & 12 & $6J $ & 6Q \\ 
\hline 0 & 6 & 12 & 10 & $10J $ & 6R \\ 
\hline 
\end{tabular} 
\end{center}
\label{TI}
\end{table}

It is quite evident that the procedure illustrated in above on relatively small regular Ising spin clusters with the total number of spins $N \leq 12$ would become rather cumbersome when applying it to much larger spin clusters as for instance the Ising dodecahedron with $N=20$. It is should be nevertheless pointed out that the main difficulty by adopting this calculation procedure for more extensive Ising spin clusters does not concern with enumerating types of induced subgraphs inherent to individual spin configurations (this task is relatively straightforward), but rather in a combinatorial assignment of respective degeneracies that must be attributed to each individual induced subgraph. However, the mathematical procedure worked out previously purely on analytic grounds is amenable for the numerical implementation as well. To illustrate the case, we have implemented a set of procedures in MAPLE computer algebra system in order to enumerate all induced subgraphs (and their respective occurrences), which are pertinent to available spin configurations of the Ising dodecahedron. For simplicity, Tab.~\ref{TD} summarizes just the lowest-energy spin configurations of the Ising dodecahedron for non-negative values of the total spin $S_T \geq 0$. It can be readily understood from Tab.~\ref{TD} that the spin configuration with eight disjunct flipped spins (the induced subgraph 8A in Fig.~\ref{fig2}) is favorable ground state of the Ising dodecahedron in a rather wide interval of the magnetic fields ranging from zero up to saturation field.   

\begin{table}
\vspace{-1.0cm}
\caption{The Ising dodecahedron (see Tab.\ref{TT} for notation). Only the lowest-energy spin configurations with $S_T \geq 0$ are listed.}
\vspace{-0.2cm}
\begin{center}
\begin{tabular}{|c|c|c|c|c|c|}
\hline $ S_{T} $ & $ n_t $ & deg & $ \tilde{d}_t $ & energy & subgraph \\  
\hline \textbf{20} & \textbf{0} & \textbf{1} & \textbf{0} & $\boldsymbol{30J-20h}$ & \textbf{0} \\ 
\hline 18 & 1 & 20 & 3 & $24J-18h$ & 1A \\ 
\hline 16 & 2 & 160 & 6 & $18J-16h$ & 2A \\ 
\hline 14 & 3 & 660 & 9 & $12J-14h $ & 3A \\ 
\hline 12 & 4 & 1510 & 12 & $6J-12h $ & 4A \\ 
\hline 10 & 5 & 1912 & 15 & $-10h$ & 5A \\ 
\hline 8 & 6 & 1240 & 18 & $-6J-8h$& 6A \\ 
\hline 6 & 7 & 320 & 21 & $-12J-6h$ & 7A \\ 
\hline \textbf{4} & \textbf{8} & \textbf{5} & \textbf{24} & $\boldsymbol{-18J-4h}$ & \textbf{8A} \\ 
\hline 2 & 9 & 840 & 23 & $-16J-2h$ & 9A \\ 
\hline 2 & 9 & 60 & 23 & $-16J-2h$ & 9B \\ 
\hline 0 & 10 & 240 & 24 & $-18J$ & 10A \\
\hline 
\end{tabular} 
\end{center}
\label{TD}
\end{table}

At this stage, the overall energy spectrum of the regular Ising polyhedra obtained within the simple graph-theoretical approach can be substituted to the definition of the partition function (\ref{pf}) in order to get full thermodynamics. For completeness, let us merely quote the final exact expressions for the partition function of the Ising tetrahedron
\begin{eqnarray}
Z_{T} \!\!\!&=&\!\!\! 6\exp\left(2\beta J\right) + 8\cosh(2\beta h) \nonumber \\
      \!\!\!&+&\!\!\! 2\exp\left(-6\beta J\right)\cosh(4\beta h),
\label{PFtetra}
\end{eqnarray}
the Ising octahedron 
\begin{eqnarray}
Z_{O} \!\!\!&=&\!\!\! 8 + 12 \exp(4\beta J) \nonumber \\
      \!\!\!&+&\!\!\! 6\cosh(2\beta h) [\exp(4\beta J) + 4] \nonumber \\
      \!\!\!&+&\!\!\! 12\exp(-4\beta J)\cosh(4\beta h) \nonumber \\
      \!\!\!&+&\!\!\! 2\exp(-12\beta J)\cosh(6\beta h), 
\label{PFokta}
\end{eqnarray}
the Ising cube
\begin{eqnarray}
Z_{C} \!\!\!&=&\!\!\! 2\exp(12\beta J) + 30\exp(4\beta J) + 32 + 6\exp(-4\beta J) \nonumber \\
      \!\!\!&+&\!\!\! 16\cosh(2\beta h) [\exp\left(6\beta J\right) + 6\cosh\left(2\beta J\right)] \nonumber \\
      \!\!\!&+&\!\!\! 8\cosh(4\beta h)[4 + 3\exp(-4\beta J)]   \nonumber \\
      \!\!\!&+&\!\!\! 16\exp\left(-6\beta J\right)\cosh(6\beta h) \nonumber \\
      \!\!\!&+&\!\!\! 2\exp(-12\beta J)\cosh(8\beta h), 
 \label{PFhexa}
\end{eqnarray}
and the Ising icosahedron
\begin{eqnarray}
Z_{I} \!\!\!&=&\!\!\! 72\exp\left(10\beta J\right) + 340\exp\left(6\beta J\right) + 300\exp\left(2\beta J\right) \nonumber \\ 
      \!\!\!&+&\!\!\! 120\exp\left(-2\beta J\right) + 80\exp\left(-6\beta J\right) + 12\exp\left(-10\beta J\right) \nonumber \\ 
      \!\!\!&+&\!\!\! 24\cosh(2\beta h) [10 \exp(8\beta J)+ 25\exp(4\beta J) + 21] \nonumber \\ 
      \!\!\!&+&\!\!\! 120\cosh(2\beta h) [\exp(-4\beta J) + \exp(-8\beta J)] \nonumber \\
      \!\!\!&+&\!\!\! 30\cosh(4\beta h) [2\exp\left(6\beta J\right) + 13\exp\left(2\beta J\right)] \nonumber \\ 
      \!\!\!&+&\!\!\! 120\cosh(4\beta h) [3\exp\left(-2\beta J\right) + \exp\left(-6\beta J\right)] \nonumber \\
      \!\!\!&+&\!\!\! 60\exp\left(-10\beta J\right)\cosh(4\beta h) \nonumber \\
      \!\!\!&+&\!\!\! 40\exp(-12\beta J)\cosh(6\beta h) \nonumber \\
      \!\!\!&+&\!\!\! 40\cosh(6\beta h) [1 + 6 \exp(-4\beta J) + 3 \exp(-8\beta J)] \nonumber \\ 
      \!\!\!&+&\!\!\! 12\cosh(8\beta h) [6 \exp\left(-10\beta J\right) + 5\exp\left(-14\beta J\right)] \nonumber \\  
      \!\!\!&+&\!\!\! 24\exp(-20\beta J)\cosh(10\beta h)  \nonumber \\   
      \!\!\!&+&\!\!\! 2\exp\left(-30\beta J\right)\cosh(12\beta h). 
\label{PFikosa}
\end{eqnarray}
Gibbs free energy normalized per one spin can be directly calculated from the partition functions of the regular Ising polyhedra according to the relation 
\begin{eqnarray}
g = -\frac{1}{N}k_{B}T\ln Z, 
\label{F}
\end{eqnarray}
which subsequently allows a straightforward calculation of the magnetization per spin  
\begin{eqnarray}
m = -\frac{\partial g}{\partial h} = \frac{1}{N} \frac{\partial \ln Z}{\partial (\beta h)} 
\label{M}
\end{eqnarray}
as well as the entropy per spin
\begin{eqnarray}
s = -\frac{\partial g}{\partial T} = \frac{k_B}{N} \left( \ln Z - \frac{1}{\beta} \frac{\partial \ln Z}{\partial \beta} \right). 
\label{S}
\end{eqnarray}
For brevity, let us merely quote the final exact analytical formula for the single-site magnetization of the Ising tetrahedron
\begin{eqnarray}
m_{T} = \frac{2}{Z_{T}}\left[2\sinh(2\beta h)+\exp\left(-6\beta J \right)\sinh(4\beta h)\right],
\label{Mtetra}
\end{eqnarray}
\newline
the Ising octahedron
\begin{eqnarray}
m_{O} \!\!\!&=&\!\!\! \frac{2}{Z_{O}} \left\{\sinh(2\beta h) [\exp(4\beta J) + 4] \right. \label{Mokta} \\ 
      \!\!\!&+&\!\!\! \left. 4\exp(-4\beta J)\sinh(4\beta h) + \exp(-12\beta J)\sinh(6\beta h) \right\}, \nonumber 
\end{eqnarray}
the Ising cube
\begin{eqnarray}
m_{C} \!\!\!&=&\!\!\! \frac{2}{Z_{C}} \left\{2\sinh(2\beta h) [\exp\left(6\beta J\right) + 6\cosh(2\beta J)] \right. \nonumber \\ 
      \!\!\!&+&\!\!\! 2\sinh(4\beta h) [3\exp(-4\beta J) + 4] \label{Mhexa} \\ 
      \!\!\!&+&\!\!\! \left. 6\exp\left(-6\beta J\right)\sinh(6\beta h) + \exp(-12\beta J)\sinh(8\beta h)  \right\}, \nonumber 
\end{eqnarray}
and the Ising icosahedron
\begin{eqnarray}
m_{I} \!\!\!&=&\!\!\! \frac{2}{Z_{I}} \left\{2\sinh(2\beta h) [10 \exp(8\beta J)+ 25\exp(4\beta J) + 21] \right.  \nonumber \\ 
      \!\!\!&+&\!\!\! 10\sinh(2\beta h) [\exp(-4\beta J) + \exp(-8\beta J)] \nonumber \\
      \!\!\!&+&\!\!\! 5\sinh(4\beta h) [2\exp\left(6\beta J\right) + 13\exp\left(2\beta J\right)] \nonumber \\ 
      \!\!\!&+&\!\!\! 20\sinh(4\beta h) [3\exp\left(-2\beta J\right) + \exp\left(-6\beta J\right)] \nonumber \\
      \!\!\!&+&\!\!\! 10\exp\left(-10\beta J\right)\sinh(4\beta h) \nonumber \\
      \!\!\!&+&\!\!\! 10\exp(-12\beta J)\sinh(6\beta h) \nonumber \\
      \!\!\!&+&\!\!\! 10\sinh(6\beta h) [1 + 6 \exp(-4\beta J) + 3 \exp(-8\beta J)] \nonumber \\ 
      \!\!\!&+&\!\!\! 4\sinh(8\beta h) [6 \exp\left(-10\beta J\right) + 5\exp\left(-14\beta J\right)] \nonumber \\  
      \!\!\!&+&\!\!\! 10\exp(-20\beta J)\sinh(10\beta h)  \nonumber \\   
      \!\!\!&+&\!\!\! \left. \exp\left(-30\beta J\right)\sinh(12\beta h) \right\}. 
\label{Mikosa}
\end{eqnarray}
It should be noted that the final expressions for the entropy of the regular Ising polyhedra obtained according to Eq. (\ref{S}) 
are too cumbersome to write them down here explicitly.  

\section{Results and discussion}
\label{result}
In this part, let us explore in detail the magnetization process and magnetocaloric properties of the regular Ising polyhedra. In particular, we will investigate a microscopic nature of highly degenerate ground states, which are manifested in low-temperature curves as profound intermediate plateaux at fractional values of the saturation magnetization.  

\begin{figure}[t]
\includegraphics[width=0.5\textwidth]{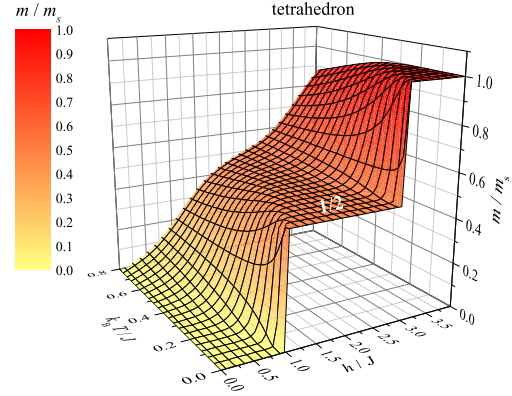}
\hspace{-1cm}
\includegraphics[width=0.5\textwidth]{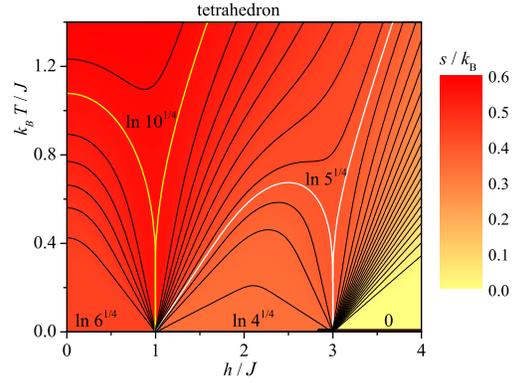}
\vspace{-1cm}
\caption{The upper panel: 3D surface plot for the magnetization of the Ising tetrahedron against the magnetic field and temperature; 
the lower panel: adiabatic temperature changes of the Ising tetrahedron upon varying the magnetic field. The magnetization is normalized with respect to its saturation value and entropy per spin.}
\label{tetr}
\end{figure}

The magnetization of the Ising tetrahedron is plotted in the upper panel of Fig.~\ref{tetr} against temperature and magnetic field. It can be clearly seen from this figure that a zero-temperature magnetization curve of the Ising tetrahedron shows two intermediate plateaux at zero and one-half of the saturation magnetization. The magnetization is zero in a field range $h/J \in (0,1)$, which favors the six-fold degenerate ground state with two spins oriented in a direction of the magnetic field and the other two spins pointing in opposite direction. One out of two spins oriented contrary to the magnetic field is flipped at the first critical field $h_{c}/J = 1$, above which the four-fold degenerate ground state emerges with three spins aligned into the magnetic field and one spin pointing in opposite direction. The aforementioned ground state manifest itself as an intermediate plateau at one-half of the saturation magnetization, which can be detected in a field range $h/J \in (1,3)$. Finally, the fully polarized state with all four spins aligned towards the magnetic field becomes the ground state above the second critical field $h_{c}/J = 3$. It should be pointed out that exact magnetization jumps and plateaux occur only at zero temperature, because a rising temperature generally causes a gradual smoothening of the magnetization curve. As a result, magnetization plateaux generally shrink and magnetization jumps are smeared out with increasing temperature.

Let us also examine magnetocaloric properties of the Ising tetrahedron, which can be particularly interesting especially in a vicinity of the aforedescribed magnetization jumps. For this purpose, the lower panel of Fig.~\ref{tetr} displays a few isentropy lines of the Ising tetrahedron in the magnetic field-temperature plane, which can be alternatively viewed as a temperature response with respect to a variation of the magnetic field under the adiabatic condition. It actually turns out that the Ising tetrahedron may exhibit an enhanced magnetocaloric effect in a vicinity of two magnetization jumps, above (below) which an abrupt rise (drop) in temperature is achieved by decreasing magnetic field whenever the entropy per spin is set close enough to the ground-state degeneracy pertinent to the first and second critical field $s/k_{\rm{B}} = \frac{1}{4} \ln 10 \doteq 0.576 $ and $S/k_{\rm{B}} = \frac{1}{4} \ln 5 \doteq 0.402$, respectively.

\begin{figure}[t]
\includegraphics[width=0.5\textwidth]{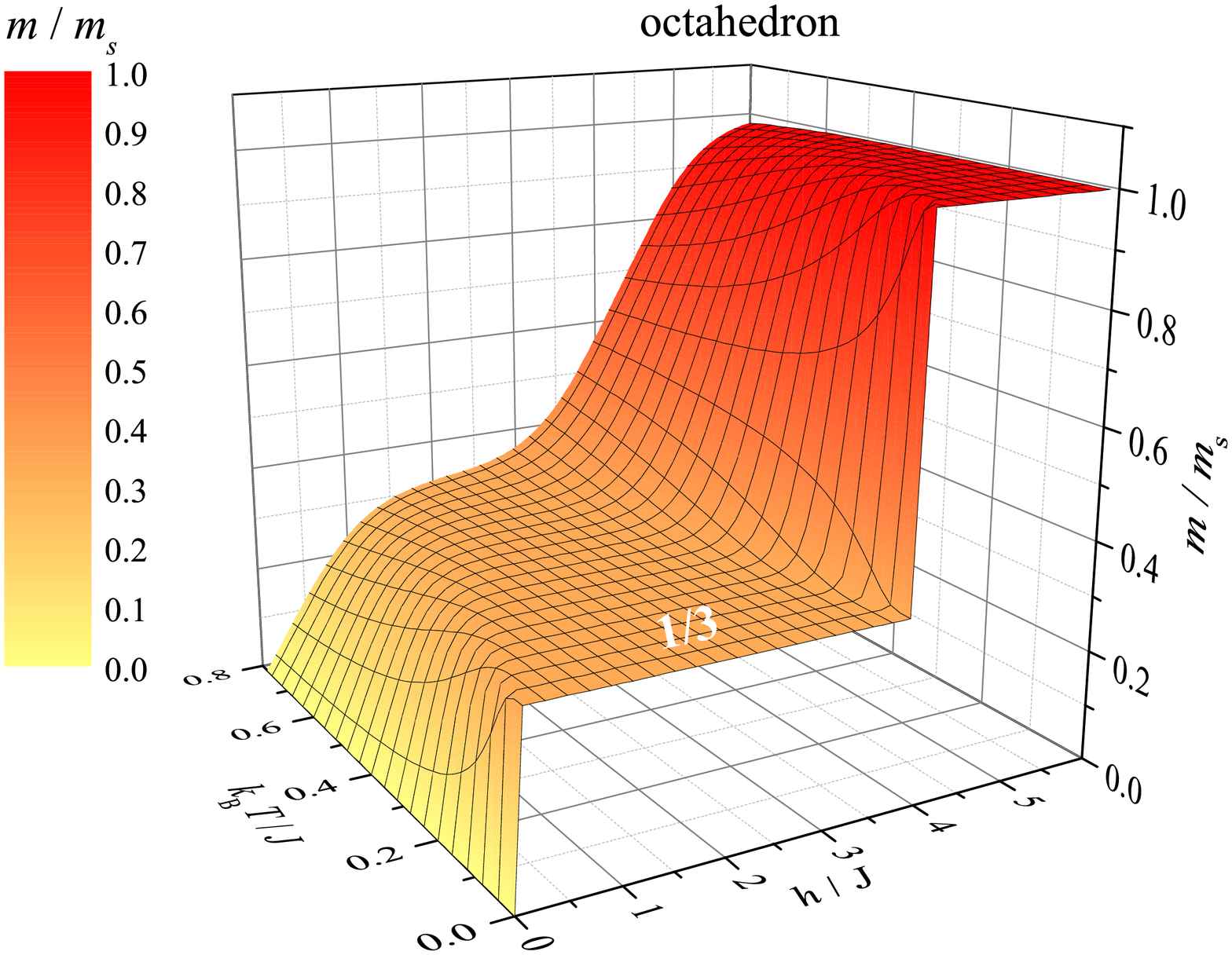}
\hspace{-1cm}
\includegraphics[width=0.5\textwidth]{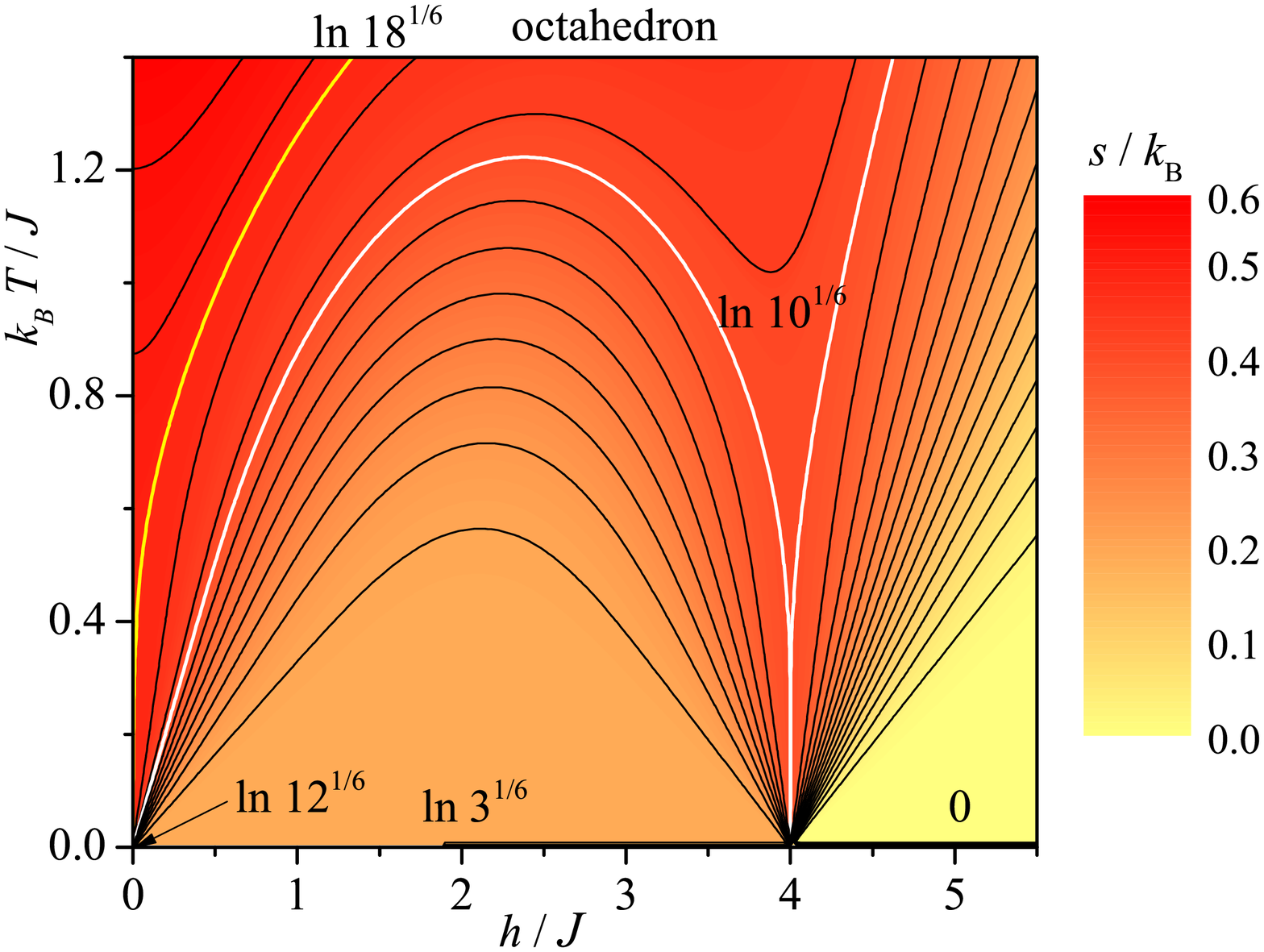}
\vspace{-1cm}
\caption{The upper panel: 3D surface plot for the magnetization of the Ising octahedron against the magnetic field and temperature; 
the lower panel: adiabatic temperature changes of the Ising octahedron upon varying the magnetic field. The magnetization is normalized with respect to its saturation value and entropy per spin.}
\label{octa}
\end{figure}

In contrast to the Ising tetrahedron, the magnetization curve of the Ising octahedron shows just one wide plateau at one-third of the saturation magnetization in a respective field range $h/J \in (0,4)$, see the upper panel in Fig.~\ref{octa}. It is worthwhile to remark that the intermediate plateau at zero magnetization is totally absent in a relevant magnetization process, because three spin configurations with two antipodal spins oriented against the magnetic field are energetically favorable before any spin configuration with the total spin $S_T=0$ as soon as the magnetic field becomes non-zero. In virtue of this fact, the zero-temperature magnetization curve of the Ising octahedron exhibits an abrupt jump from zero to one-third of the saturation magnetization right at zero field, whereas the intermediate one-third plateau terminates at the saturation field $h_{c}/J = 4$ that determines a field-induced transition towards the fully saturated state. The effect of temperature upon a gradual smoothening of the magnetization curve is quite similar as in the case of the Ising tetrahedron.

Adiabatic changes of temperature achieved upon varying the magnetic field are depicted for the Ising octahedron in the lower panel of Fig.~\ref{octa}. Apparently, the Ising octahedron shows an enhanced magnetocaloric effect near the saturation field provided that the entropy per spin is close to the ground-state degeneracy relevant for the saturation field $s/k_{\rm{B}} = \frac{1}{6} \ln 10 \doteq 0.384$. However, it is worth noticing that an absence of intermediate plateau at zero magnetization has significant impact on magnetocaloric properties of the Ising octahedron. The absence of zero magnetization plateau bears a close relation to a giant magnetocaloric effect observable near zero magnetic field. If the entropy is selected close enough to the ground-state degeneracy emergent at zero field $s/k_{\rm{B}} = \frac{1}{6} \ln 18 \doteq 0.482$, then, the adiabatic demagnetization of the Ising octahedron is characterized by a rapid decrease in temperature as the magnetic field diminishes. It should be stressed that the isentropy line at $s/k_{\rm{B}} = \frac{1}{6} \ln 18 \doteq 0.482$ reaches the absolute zero temperature with an infinite gradient as the magnetic field tends to zero, which suggest the fastest possible cooling process that would be quite superior with respect to that of paramagnetic salts. A refrigeration with the help of adiabatic demagnetization of the Ising octahedron might be therefore of substantial technological importance, since the efficient cooling persists in a rather wide range of the magnetic fields and the ultra-low temperatures can be consequently reached quite efficiently. 

\begin{figure}[t]
\includegraphics[width=0.5\textwidth]{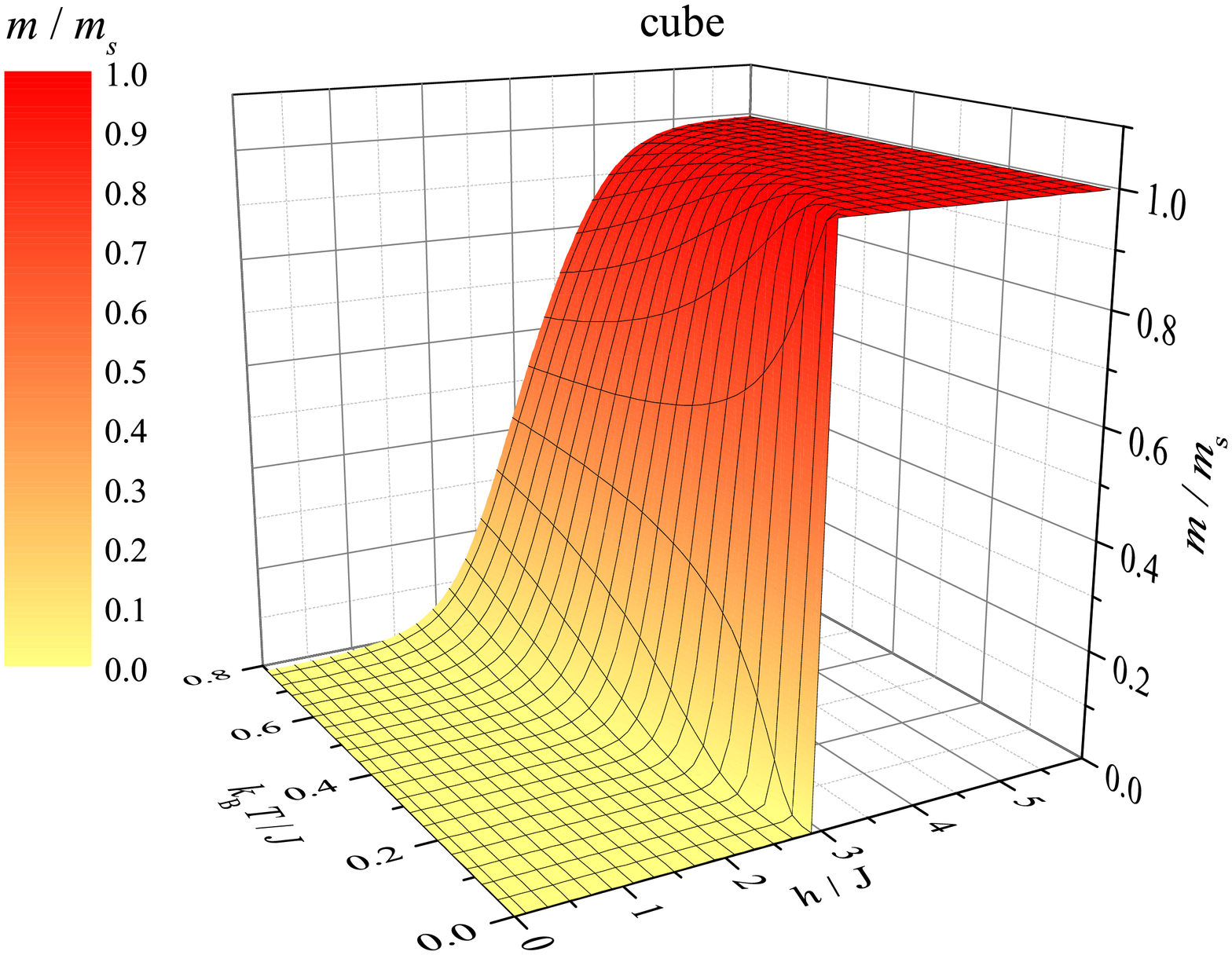}
\hspace{-1cm}
\includegraphics[width=0.5\textwidth]{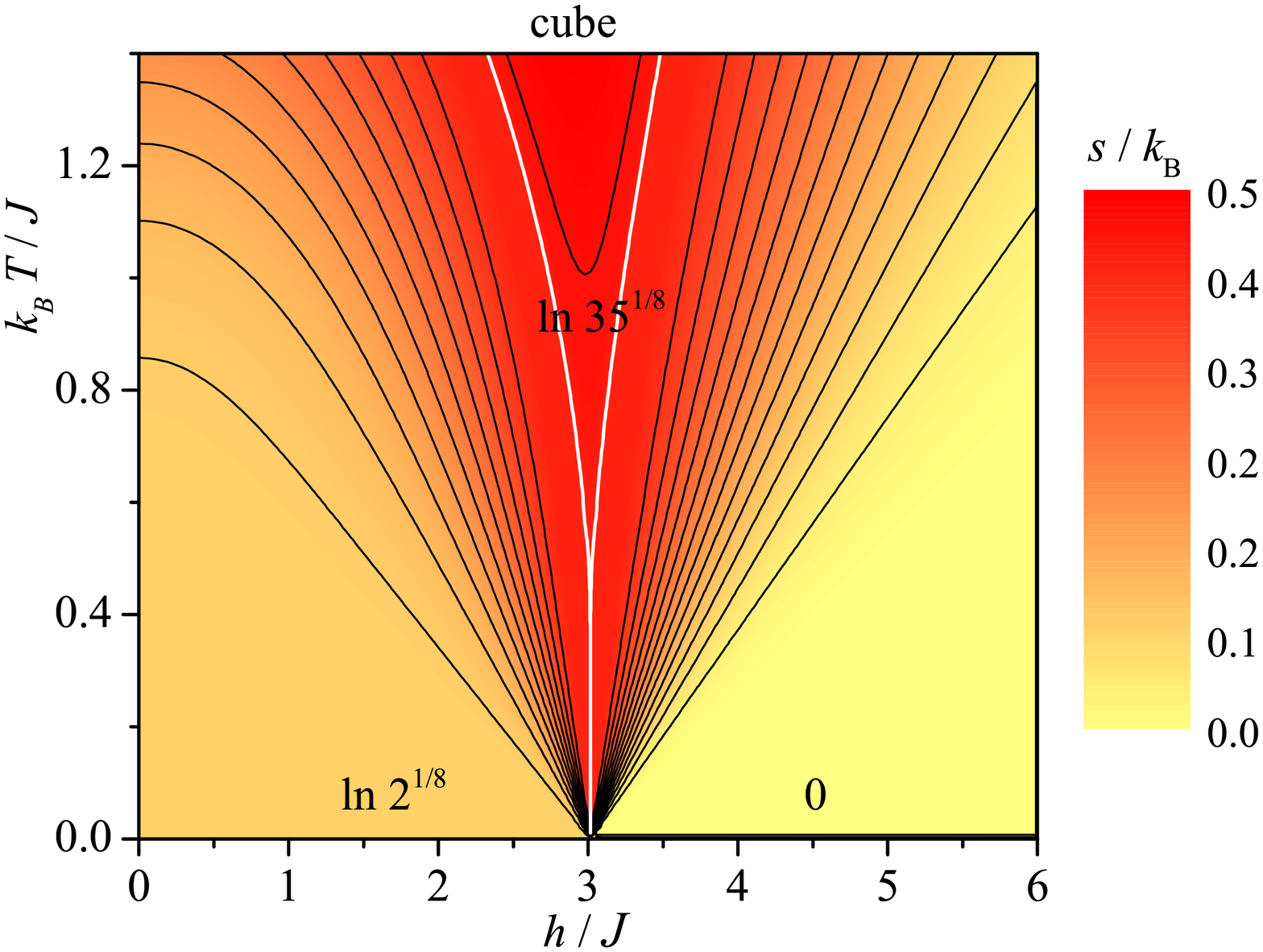}
\vspace{-1cm}
\caption{The upper panel: 3D surface plot for the magnetization of the Ising cube against the magnetic field and temperature; 
the lower panel: adiabatic temperature changes of the Ising cube upon varying the magnetic field. The magnetization is normalized with respect to its saturation value and entropy per spin.}
\label{cube}
\end{figure}

The magnetization curve of the Ising cube as the only non-frustrated regular polyhedron consists of a single plateau at zero magnetization, which is subsequently followed by a direct magnetization jump from zero to the saturation magnetization (see the upper panel in Fig. \ref{cube}). The bipartite nature of the Ising cube enables an existence of two-fold degenerate antiferromagnetic ground state at sufficiently low magnetic fields with four reversed spins, which undergo a simultaneous spin flip to the fully polarized state at the saturation field $h_{c}/J = 3$. It is noteworthy that the magnetization curve becomes smoother upon increasing temperature and there is a huge degeneracy precisely at the saturation field, at which all lowest-energy spin configurations with $S_T \geq 0$ coexist together. It could be concluded that the vertex-transitive bipartite Ising spin clusters cannot exhibit any other magnetization plateau except that at zero magnetization.

The adiabatic demagnetization of the Ising cube is displayed in the lower panel of Fig. \ref{cube}, which shows a contour plot of entropy as a function of temperature and magnetic field. The giant magnetocaloric effect is observed in a vicinity of the saturation field on account of a huge degeneracy of the lowest-energy spin configurations with $S_T \geq 0$ on assumption that the entropy per spin is set to $s/k_{\rm{B}} = \frac{1}{12} \ln35 \doteq 0.296$. However, 
a  rather steep decrease in temperature observable slightly above the saturation field is successively followed by a subsequent steep increase in temperature below the saturation field, which consequently precludes a possible utilization of a giant magnetocaloric effect for cooling purposes. 

\begin{figure}[t]
\includegraphics[width=0.5\textwidth]{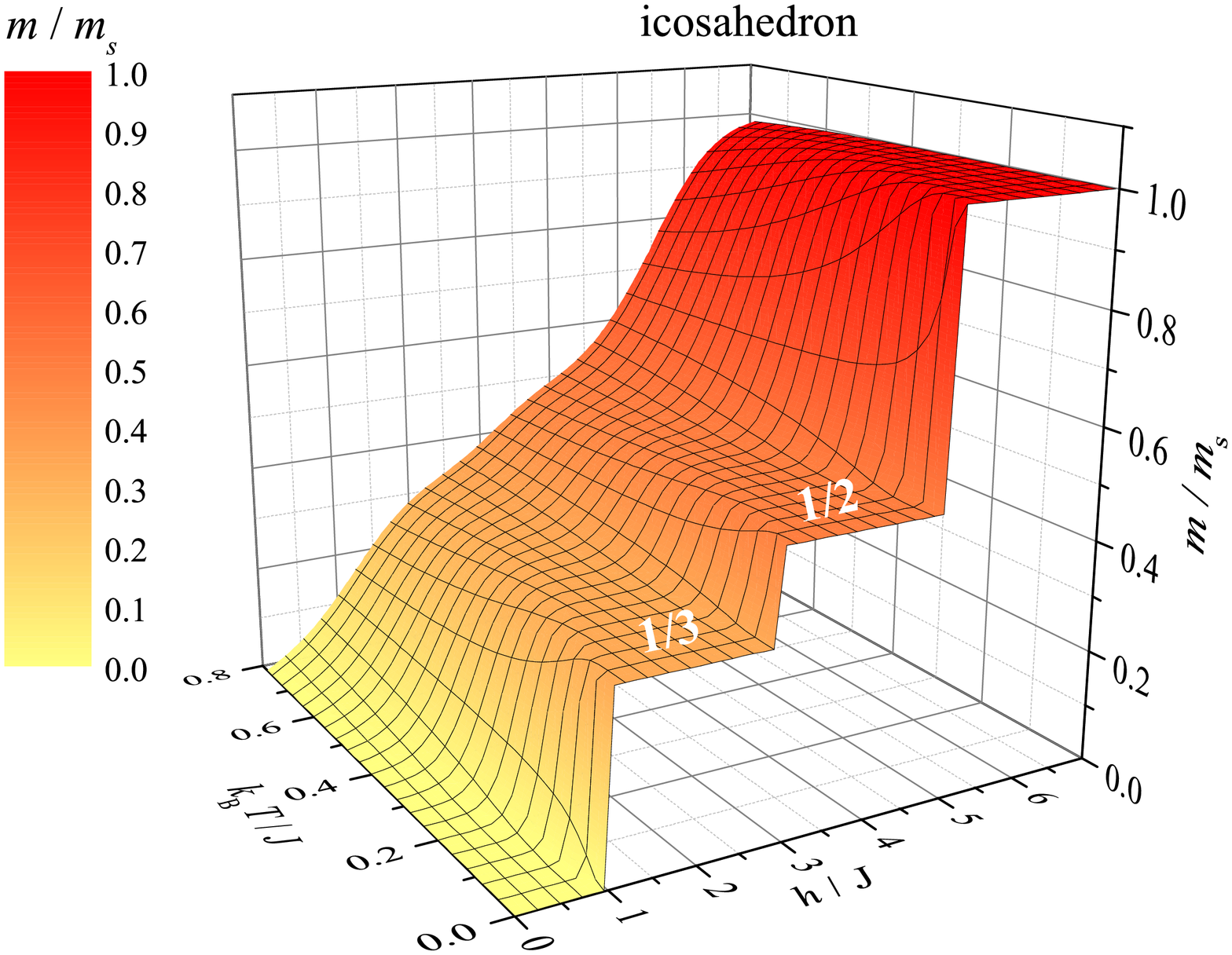}
\hspace{-1cm}
\includegraphics[width=0.5\textwidth]{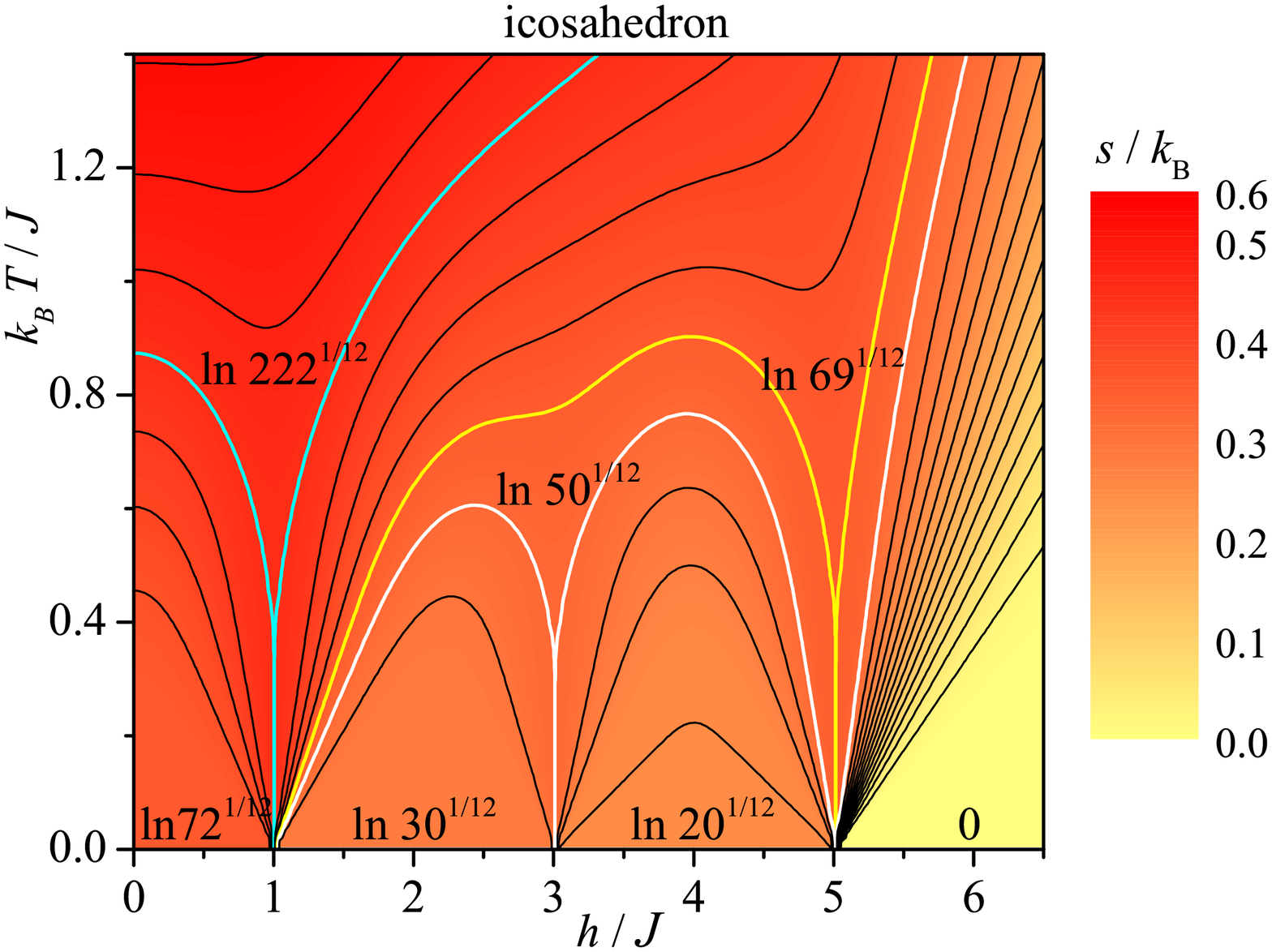}
\vspace{-1cm}
\caption{The upper panel: 3D surface plot for the magnetization of the Ising icosahedron against the magnetic field and temperature; 
the lower panel: adiabatic temperature changes of the Ising icosahedron upon varying the magnetic field. The magnetization is normalized with respect to its saturation value and entropy per spin.}
\label{icos}
\end{figure}

The most remarkable magnetization process evidently exhibits the Ising icosahedron, which displays in total three different intermediate plateaux at zero, one-third and one-half of the saturation magnetization (see the upper panel in Fig. \ref{icos}). At sufficiently low magnetic fields $h/J \in (0,1)$ the ground state is highly (72-fold) degenerate due to an accidental degeneracy of two lowest-energy spin configurations, in which six reversed spins are located at vertices that either form an open chain of six sites or a closed chain of five sites plus one additional isolated site (see the corresponding induced subgraphs 6B and 6C in Fig. \ref{fig2}). The intermediate one-third magnetization plateau develops at moderate fields $h/J \in (1, 3)$ on account of the another highly (30-fold) degenerate ground state, in which two out of four reversed spins are located at isolated vertices and the other two inverted spins at adjacent vertices (see the corresponding induced subgraph 4B in Fig. \ref{fig2}). The last intermediate plateau at one-half of the saturation magnetization occurs in the field range $h/J \in (3, 5)$, which prefers the highly (20-fold) degenerate ground state with three isolated spins oriented in opposite to the magnetic field (see the corresponding induced subgraph 3A in Fig. \ref{fig2}). Apparently, the high degeneracy of all aforedescribed ground states is closely connected to a geometric spin frustration of the Ising icosahedron, which is progressively lowered with an increase of the external magnetic field until the degeneracy is completely lifted above the saturation field $h_{c}/J = 5$ due to the unique fully ordered spin configuration. 

To achieve an enhanced magnetocaloric effect during the adiabatic demagnetization of the Ising icosahedron (the lower panel in Fig. \ref{icos}), it is necessary to set entropy close enough to an appropriate ground-state degeneracy at a critical field related to the respective magnetization jump. Thus, a gradual decrease in the magnetic field is responsible for the fast cooling (heating) slightly above (below) the relevant critical field only if entropy per spin is close to $s/k_{\rm{B}} = \frac{1}{12} \ln222 \doteq  0.450$ nearby the first critical field $h_{c}/J = 1$, $s/k_{\rm{B}} = \frac{1}{12} \ln50 \doteq  0.326$ nearby the second critical field $h_{c}/J = 3$ and $s/k_{\rm{B}} = \frac{1}{12} \ln69 \doteq 0.353$ nearby the saturation field $h_{c}/J = 5$.  

\begin{figure}[t]
\includegraphics[width=0.5\textwidth]{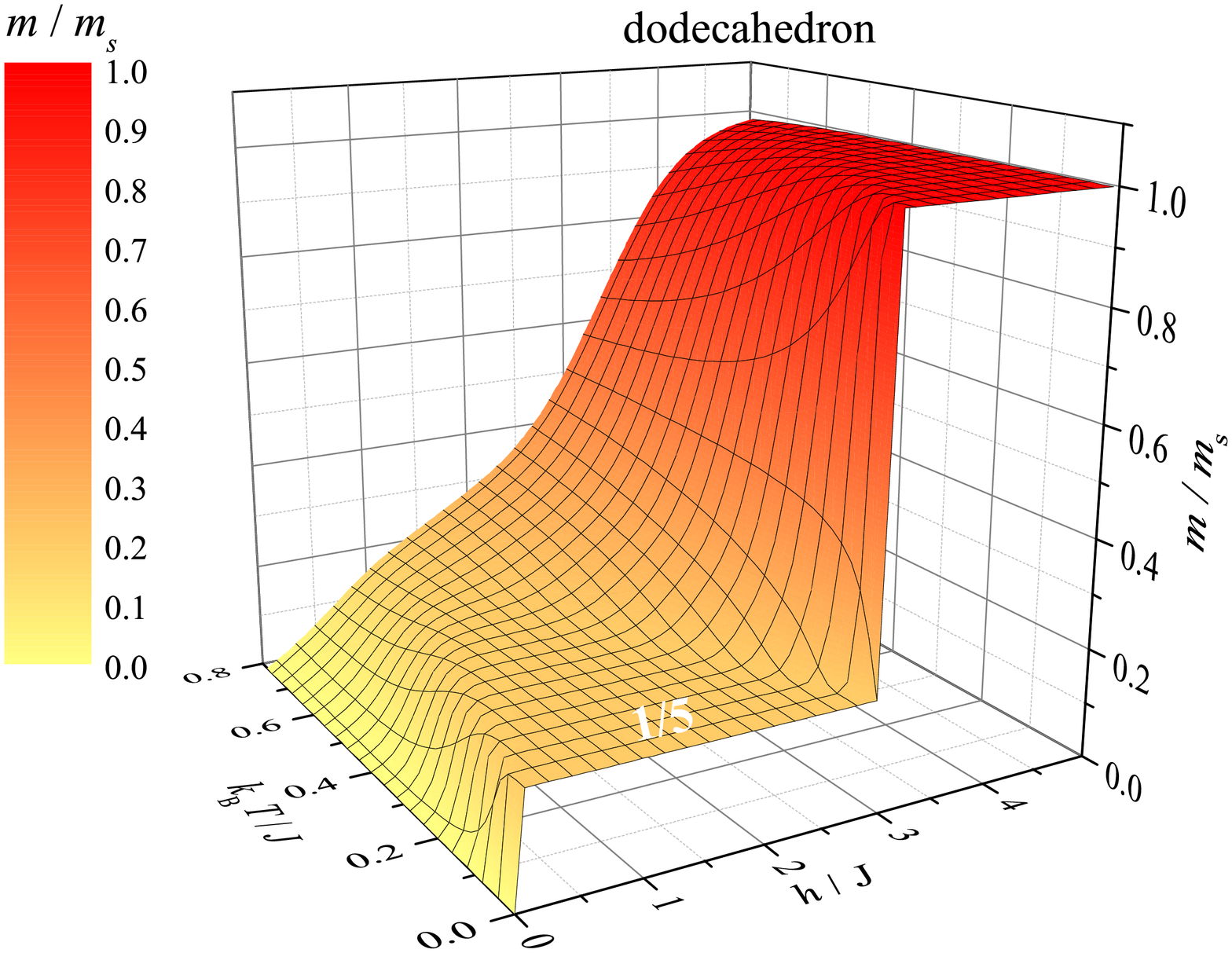}
\hspace{-1cm}
\includegraphics[width=0.5\textwidth]{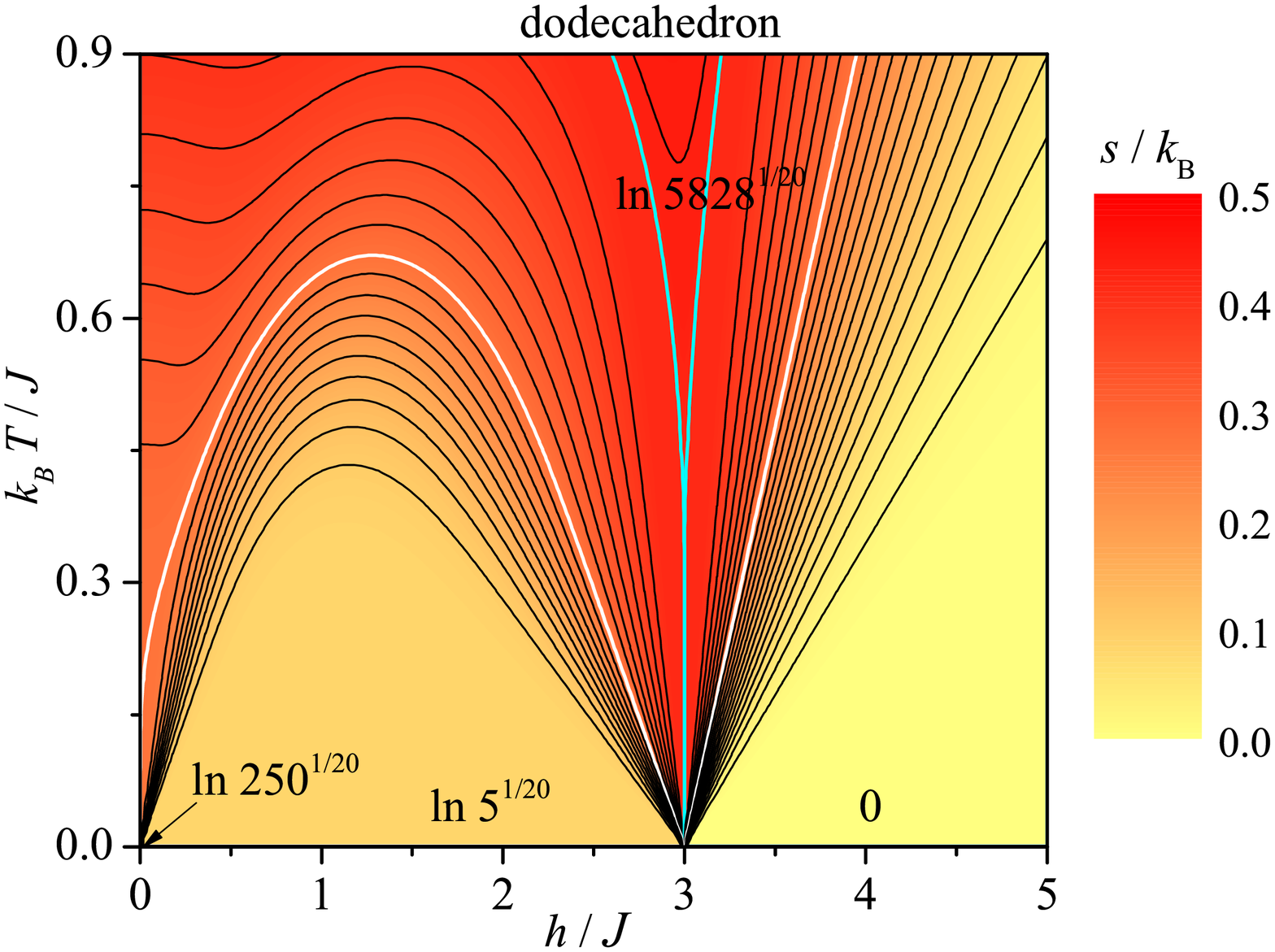}
\vspace{-1cm}
\caption{The upper panel: 3D surface plot for the magnetization of the Ising dodecahedron against the magnetic field and temperature; 
the lower panel: adiabatic temperature changes of the Ising dodecahedron upon varying the magnetic field. The magnetization is normalized with respect to its saturation value and entropy per spin.}
\label{dode}
\end{figure}

The magnetization process of the Ising dodecahedron (the upper panel in Fig. \ref{dode}) is quite reminiscent of that of the Ising octahedron. The zero-temperature magnetization curve of the Ising dodecahedron shows an abrupt jump at zero magnetic field, which is consecutively followed by a rather wide intermediate plateau at one-fifth of the saturation magnetization ending up only at the saturation field $h_{c}/J = 3$. The ground state of the Ising dodecahedron is just five-fold degenerate despite of a strong spin frustration in a low-field region $h/J \in (0, 3)$, where five different spin configurations with eight isolated spins aligned contrary to the magnetic field become the lowest-energy microstates. The magnetization plateau and jump are gradually smoothened with an increase in temperature until they completely vanish above a certain temperature. 

One of the most important features of the magnetization process of the Ising dodecahedron is the absence of zero magnetization plateau, which results from an incapability of microstates with the total spin $S_T = 0$ or $1$ to further reduce the configurational energy of the ground-state manifold composed of five microstates with the total spin $S_T=2$. Owing to this fact, the giant magnetocaloric effect can be detected during the adiabatic demagnetization of the Ising dodecahedron in a very low-field region on assumption that entropy per spin is close enough to the ground-state degeneracy at zero field $s/k_{\rm{B}} = \frac{1}{20} \ln250 \doteq 0.276$. From this perspective, the Ising dodecahedron could be regarded together with the Ising octahedron as the most prominent frustrated Ising spin cluster that would allow efficient cooling through the adiabatic demagnetization.

\section{Conclusion}
\label{conclusion}

The present work deals with the magnetization process and magnetocaloric properties of the regular Ising polyhedra (tetrahedron, octahedron, cube, icosahedron and dodecahedron), which are rigorously calculated by means of a relatively simple graph-theoretical approach. Apart from the exact analytic calculations, the proposed procedure is suitable also for the numerical implementation as it was illustrated on the particular example of the Ising dodecahedron. In addition, the method can be rather straightforwardly adapted in order to calculate magnetic properties of the irregular Ising spin clusters with a more complex geometric structure as well.

It has been demonstrated that the magnetic behaviour of antiferromagnetic Ising spin clusters with the shape of regular polyhedra is basically influenced through a geometric spin frustration with exception of the Ising cube, which is the only bipartite regular polyhedron. In a consequence of that, the magnetization curve of the Ising cube involves just one trivial plateau at zero magnetization due to a simple antiferromagnetic ground state in contrast to the magnetization curves of all other regular Ising polyhedra. Among these, the Ising tetrahedron and icosahedron still exhibit a zero magnetization plateau, which is successively followed by the other intermediate one-half plateau in the former case and respectively, the intermediate one-third and one-half plateaux in the latter case. On the other hand, there is a lack of zero magnetization plateau in the magnetization curves of the Ising octahedron and dodecahedron, which display only a single intermediate plateau at one-third and one-fifth of the saturation magnetization, respectively.  

The intermediate magnetization plateaux emerging in a low-temperature magnetization process owing to a geometric spin frustration reveal highly degenerate ground states, the microscopic nature of which was clarified along with their respective degeneracies. We have found convincing evidence that the higher number of magnetization plateaux does not necessarily imply a quantum nature of higher-order magnetization plateaux contrary to conventional expectations \cite{yama00,saka00}, but it may be just a simple consequence of geometric spin frustration. It has been also evidenced that the magnetization jumps accompanying a transition between different magnetization plateaux manifest themselves in a giant magnetocaloric effect during the adiabatic demagnetization. From this point of view, the Ising octahedron and dodecahedron have turned out as the most prominent geometrically frustrated spin clusters that would allow an efficient cooling down to lowest achievable temperatures.

\end{document}